\documentclass[aps,twocolumn,superscriptaddress,showkeys]{revtex4}
\usepackage{times}
\usepackage{graphics}
\usepackage[dvips]{graphicx}
\usepackage{multirow}

\begin{document}

\title{Statistical physics of crime: A review}

\author{Maria R. D'Orsogna}
\email{dorsogna@csun.edu}
\affiliation{Department of Mathematics, California State University at Northridge, Los Angeles, CA 91330, USA}
\affiliation{Department of Biomathematics, UCLA, Los Angeles, CA 90095, USA}

\author{Matja\v{z} Perc}
\email{matjaz.perc@um.si}
\affiliation{Faculty of Natural Sciences and Mathematics, University of Maribor, Koro{\v s}ka cesta 160, SI-2000 Maribor, Slovenia}
\affiliation{Department of Physics, Faculty of Science, King Abdulaziz University, Jeddah, Saudi Arabia}
\affiliation{CAMTP -- Center for Applied Mathematics and Theoretical Physics, University of Maribor, Krekova 2, SI-2000 Maribor, Slovenia}

\begin{abstract}
Containing the spreading of crime in urban societies remains a major challenge. Empirical evidence suggests that, left unchecked, crimes may be recurrent and proliferate. On the other hand, eradicating a culture of crime may be difficult, especially under extreme social circumstances that impair the creation of a shared sense of social responsibility. Although our understanding of the mechanisms that drive the emergence and diffusion of crime is still incomplete, recent research highlights applied mathematics and methods of statistical physics as valuable theoretical resources that may help us better understand criminal activity. We review different approaches aimed at modeling and improving our understanding of crime, focusing on the nucleation of crime hotspots using partial differential equations, self-exciting point process and agent-based modeling, adversarial evolutionary games, and the network science behind the formation of gangs and large-scale organized crime. We emphasize that statistical physics of crime can relevantly inform the design of successful crime prevention strategies, as well as improve the accuracy of expectations about how different policing interventions should impact malicious human activity deviating from social norms. We also outline possible directions for future research, related to the effects of social and coevolving networks and to the hierarchical growth of criminal structures due to self-organization.
\end{abstract}


\maketitle

\section{Introduction}
\label{intro}
An unattended broken window invites bypassers to behave mischievously or even disorderly. Soon, one broken window may become many, and the inception of urban decay is in place. Similarly, a subway graffiti, however beautiful and harmless in appearance, points to an unkept environment that anyone can desecrate, signaling that more egregious damage will be tolerated. Panhandlers, drunks, addicts, prostitutes, and loiterers are more likely to frequent neglected subway stations than orderly and carefully patrolled ones. The 1982 seminal paper by Wilson and Kelling \cite{WILSON_AM82} contains many more lucid examples and anecdotes to introduce the ``broken windows theory'', articulating how seemingly unimportant and petty signals of urban disorder may elicit antisocial behavior and serious crime. Although not immune from criticism, this work has since become a widely adopted criminological theory.

\begin{figure}
\centering{\includegraphics[width = 8.5cm]{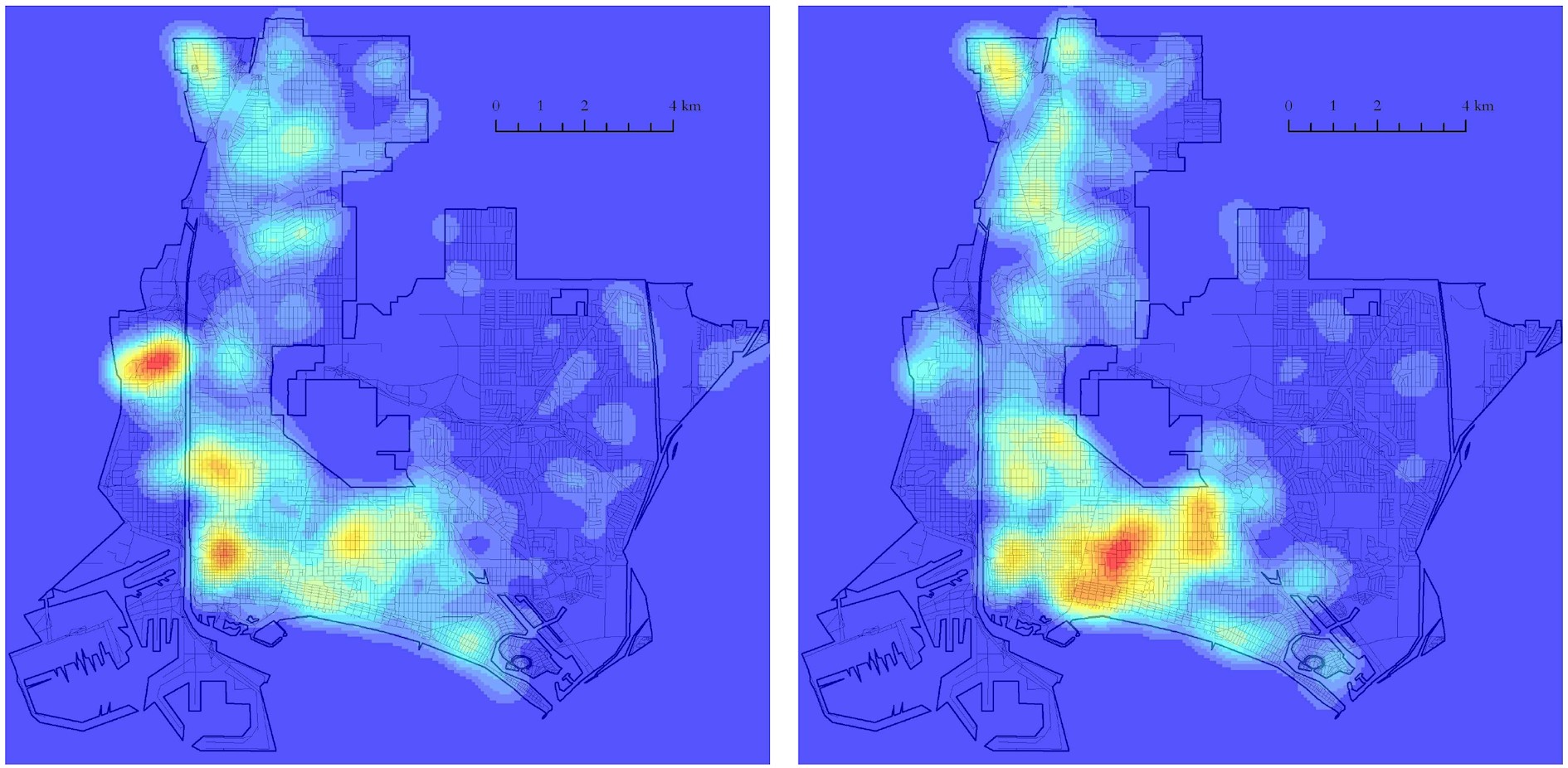}}
\caption{Dynamic changes in residential burglary hotspots for two consecutive three-month periods, starting June 2011, in Long Beach, California. The emergence of different burglary patterns is related to how offenders move within their environments and how they respond to the successes and failures of their illicit activities.  Residential burglars tend to return to previously victimized locations, or to their close vicinities, after having acquired information on the properties, the schedules of inhabitants, possible surveillance systems -- a reasoning that is closely aligned with ``routine activity theory'' \cite{COHEN_ASR79}. The figure is reproduced from \cite{SHORT_MMMAS08}.}
\label{hot}
\end{figure}

To mathematicians and physicists, the broken windows theory may be reminiscent of complexity science and self-organized criticality \cite{BAK_96}, where seemingly small and irrelevant changes at the local level frequently have unexpected consequences at the global level later in time. Feedback loops, bifurcations and catastrophes \cite{KUZNETSOV_95}, as well as phase transitions \cite{STANLEY_71}, are commonly associated with emergent phenomena stemming from the nonlinearities inherent to complex social systems \cite{CASTELLANO_RMP09}. Crime is ubiquitous, yet far from being uniformly distributed across space or time \cite{BRANTINGHAM_84, CHAINEY_05, FELSON_06, ALVES_PA13, ALVES_PONE13, PICOLI_SR14}. This is evidenced also by the dynamic nucleation and dissipation of crime hotspots shown in Fig.~\ref{hot} \cite{SHORT_MMMAS08, SHORT_PNAS10}, as well as by the emergence of complex geographical gang and organized crime networks. Such intriguing pattern formation naturally invites quantitative mathematical analyses, to which we attend in this review.

\begin{figure}
\centering{\includegraphics[width = 8.5cm]{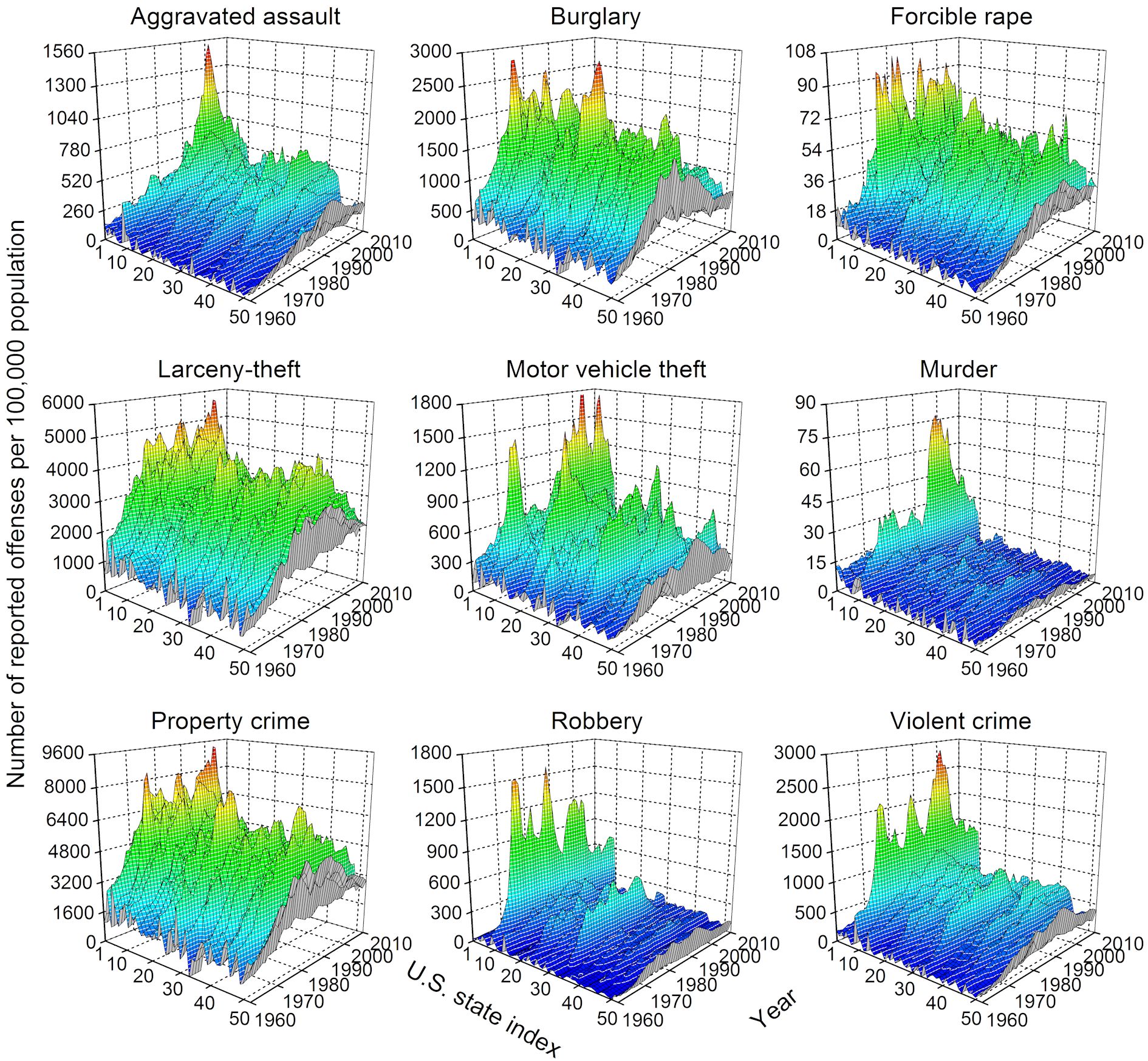}}
\caption{The persistence of crime over time despite our best prevention and punishment efforts. Data from the Federal Bureau of Investigation indicate that crime, regardless of type and severity, is remarkably recurrent. Although positive and negative trends may be inferred, crime events (measured as number of offenses per 100,000 population) between 1960 and 2010 fluctuate across time and space. There is no evidence to support that crime rates are permanently decreasing. The U.S. state index is alphabetical, including the District of Columbia being 9th, and the U.S. total being 52th.}
\label{data}
\end{figure}

We consider crime as a complex phenomenon, where nonlinear feedback loops and self-organization give rise to system-wide unexpected behaviors that are difficult to understand and control \cite{BALL_12}. Data provided by the Federal Bureau of Investigation shown in Fig.~\ref{data} suggest that crime deterrence policies are struggling to have the desired impact. Indeed, if viewed over a time scale of decades, the relative frequency of offenses, regardless of crime type,  is heavily undulating and lacks persistent downward momentum.

Outside the realm of mathematical modeling, there exist well-known and widely accepted theories of criminal behavior. According to ``routine activity theory'' \cite{COHEN_ASR79}, most criminal acts are born out of the convergence of three factors: the presence of likely offenders and of suitable targets and the absence of guardians to protect against the attempted crime. Residential burglary, grand theft auto, armed robberies, pickpocketing and rape are examples of such criminal acts. Other crimes may imply a precise target focus, such as in murder for revenge or other clan-type retaliation offenses.

If viewed upon sociologically, these ``ingredients'' of routine activity theory are relatively straightforward conditions that obviously favor criminal activity. Mathematically, however, routine activity theory allows us to model criminal offender dynamics as deviations from simple random walks. This is due to built-in heterogeneities in target selection that may drive criminal activity towards preferred locations and away from less desired ones, marked, say, by better surveillance systems, more physical obstacles to break-ins or thefts, and in general by a higher, real or perceived, risk vs. reward ratio. The degree of target ``attractiveness'' may change in time and depend on mundane factors such as the day of the week or weather conditions, or on the more sophisticated interplay between landscape, criminal activity and law enforcement responses. Crime dynamics may also include learning mechanisms or feedback loops. All these elements lead to the emergence of non-trivial patterns such as spatially localized crime hotspots as shown in Fig.~\ref{hot} for grand-theft auto in the city of Long Beach, California: note how hotspots diffuse over time in a nontrivial manner.  Another interesting phenomenon is that of repeat and near-repeat victimization in the case of residential burglary \cite{JOHNSON_BJC97, TOWNSLEY_ANZJC00, JOHNSON_BJC04, SHORT_JQC09}, whereby for a given period of time after a first break-in, the odds of a second victimization are greatly enhanced both for the original target site and locations in its near vicinity.

The complexity of crime dynamics and the many factors that influence criminal activity render mitigation and displacement of crime a non trivial task \cite{GREEN_JQ95, BRAGA_AN01, BRAGA_JEC05, WEISBURD_C06, TANIGUCHI_JQ09}. Rational choice theories applied to crime may be too simplistic in assuming straightforward gain-loss principles, for example that stronger punishment would automatically lead to less crime \cite{BECKER_JPE68, DOOB_CJ03}. In this work, we review recent quantitative mathematical models of crime where statistical physics, complexity science, game theory and self-organized criticality are used in an attempt to understand the multiple aspects of crime and to identify possible prevention and amelioration strategies.

The organization of this review is as follows. In Section~\ref{hotspots}, we will focus on a set of reaction-diffusion partial differential equations to study the emergence, dynamics and possible suppression of crime hotspots \cite{SHORT_MMMAS08, SHORT_JQC09, SHORT_SIAM10, SHORT_PNAS10, RODRIGUEZ_MMMAS10}. In Section~\ref{point}, we will review the application of self-exciting point processes, which are frequently used by seismologists to study space-time clustering of earthquakes \cite{DALEY_03} to crime data \cite{MOHLER_JASA11, MOHLER_SIAM12, LEWIS_SJ12}. Section~\ref{social} will be devoted to the study of crime by means of adversarial games and evolutionary social dilemmas \cite{SHORT_PRE10, DORSOGNA_PONE13, SHORT_EJAM13, PERC_PONE13}. In Section~\ref{gangs}, we will review mechanisms behind the growth and structure of criminal networks and the formation of gangs \cite{BARBARO_PA13}, while in Section~\ref{rehab}, we will survey a rehabilitation and recidivism model \cite{BERENJI_PONE14} that reveals an optimal resolution of the ``stick versus carrot'' dilemma \cite{SIGMUND_PNAS01, SIGMUND_TREE07, DREBER_N08, RAND_S09, HILBE_PRSB10, RAND_NC11, SZOLNOKI_NJP12, SZOLNOKI_PRE13}. We will conclude with a summary and an outlook in Section~\ref{sumlook}, describing the implication of statistical physics of crime and outlining viable directions for future research related to agent-based modeling, hierarchical growth, and self-organization.

\section{Crime hotspots}
\label{hotspots}
Empirical observations of spatiotemporal clusters of crime in urban areas, such as those presented in Fig.~\ref{hot}, motivated the development of a statistical model of criminal behavior, which was introduced and studied in \cite{SHORT_MMMAS08}. The model was developed to study residential burglary, one of the simplest instances of opportunistic, routine activity crime. Here, target sites -- residential homes -- are stationary and one can focus on burglar dynamics only, as opposed to crimes where both offender and targets are mobile, as in the case of assault or pickpocketing.

Burglars most often opportunistically victimize areas that are sufficiently, but not too close to where they live, and where they have committed crimes before \cite{JOHNSON_JQC07}. Aside from a small exclusion zone centered around their own residences, the distances that criminals are willing to travel to engage in criminal acts are best described by monotonically decreasing functions \cite{RENGERT_C99}. Offender movement is usually described as a biased random walk. The bias is twofold. One one hand, from a burglar's perspective, a given target home may be intrinsically more attractive than another due to its perceived wealth, the ease in accessing it, or the predictable routine of its occupants. On the other, there may be learned elements that bias the burglar towards a specific location. For instance, a criminal may return to a  previously victimized home or to its immediate neighborhood, having already learnt the details of the area, and having acquired the know-how for a successful break-in.

To quantify the bias towards any given location and to determine the subsequent rate of burglary, the crime model of \cite{SHORT_MMMAS08} includes a dynamically changing ``attractiveness'' field. The tendency for repeat victimization \cite{JOHNSON_BJC97, TOWNSLEY_ANZJC00, JOHNSON_BJC04, SHORT_JQC09} is included in the model by temporarily increasing the attractiveness field in response to past burglary events \cite{BERNASCO_C03, BERNASCO_BJC05} both at the burglary site and in its near vicinity. Since potential crime targets in case of residential burglary do not move, it is convenient to start with a discrete model on a square lattice with periodic boundary conditions. Each lattice site $s=(i,j)$ is a house or real estate with attractiveness $A_s(t)$. The higher the value of $A_s(t)$, the higher the bias towards site $s$ and the more likely will it be victimized. Moreover, once site $s$ has been victimized, its attractiveness further increases.  The following decomposition is introduced
\begin{equation}
A_s(t)=A^{0}_s+B_s(t),
\label{field}
\end{equation}
where $A^{0}_s$ is the static, though possibly spatially varying, component of the attractiveness field, and $B_s(t)$ represents the dynamic component associated with repeat and near-repeat victimization \cite{JOHNSON_BJC97, JOHNSON_JQC07}. Specifically, $B_s(t+1)=B_s(t)(1-\omega)+E_s(t)$, where $\omega$ sets a time scale over which repeat victimizations are most likely to occur and $E_s(t)$ is the number of burglary events that occurred at site $s$ during $t$ and $t+1$. To take into account the broken windows theory \cite{WILSON_AM82}, we let $B_s(t)$ spread locally from each site $s$ towards its nearest neighbors $s'$ according to
\begin{equation}
B_s(t+1)=\left[(1-\eta)B_s(t)+\frac{\eta}{z}\sum\limits_{s'}B_{s'}(t)\right](1-\omega)+E_s(t)
\label{discreteB}
\end{equation}
where the sum runs over the nearest neighbor sites associated to site $s$, $z$ is the coordination number of the lattice and $\eta$ is a parameter between zero and one that determines the significance of neighborhood effects. Higher values of $\eta$ lead to a greater degree of spreading of the attractiveness generated by a given burglary event, and vice-versa for lower values. In this review we assume, for simplicity, that the spacing between sites $\ell$ and the discrete time unit $\delta t$ over which criminal actions occur are both equal to one, and that every time a site $s$ is burglarized its dynamic attractiveness $B_s(t)$ increases by one. Interaction networks other than the square lattice, that may better describe the city grid or social networks for other types of crime \cite{WASSERMAN_94, ALBERT_RMP01, CHRISTAKIS_09}, and even coevolving or dynamically changing networks \cite{SANTOS_PLOSCB06, VAN-SEGBROECK_PRL09, PERC_BS10, MOREIRA_SREP13}, can be easily accommodated in a similar way.

\begin{figure}
\centering{\includegraphics[width = 8.5cm]{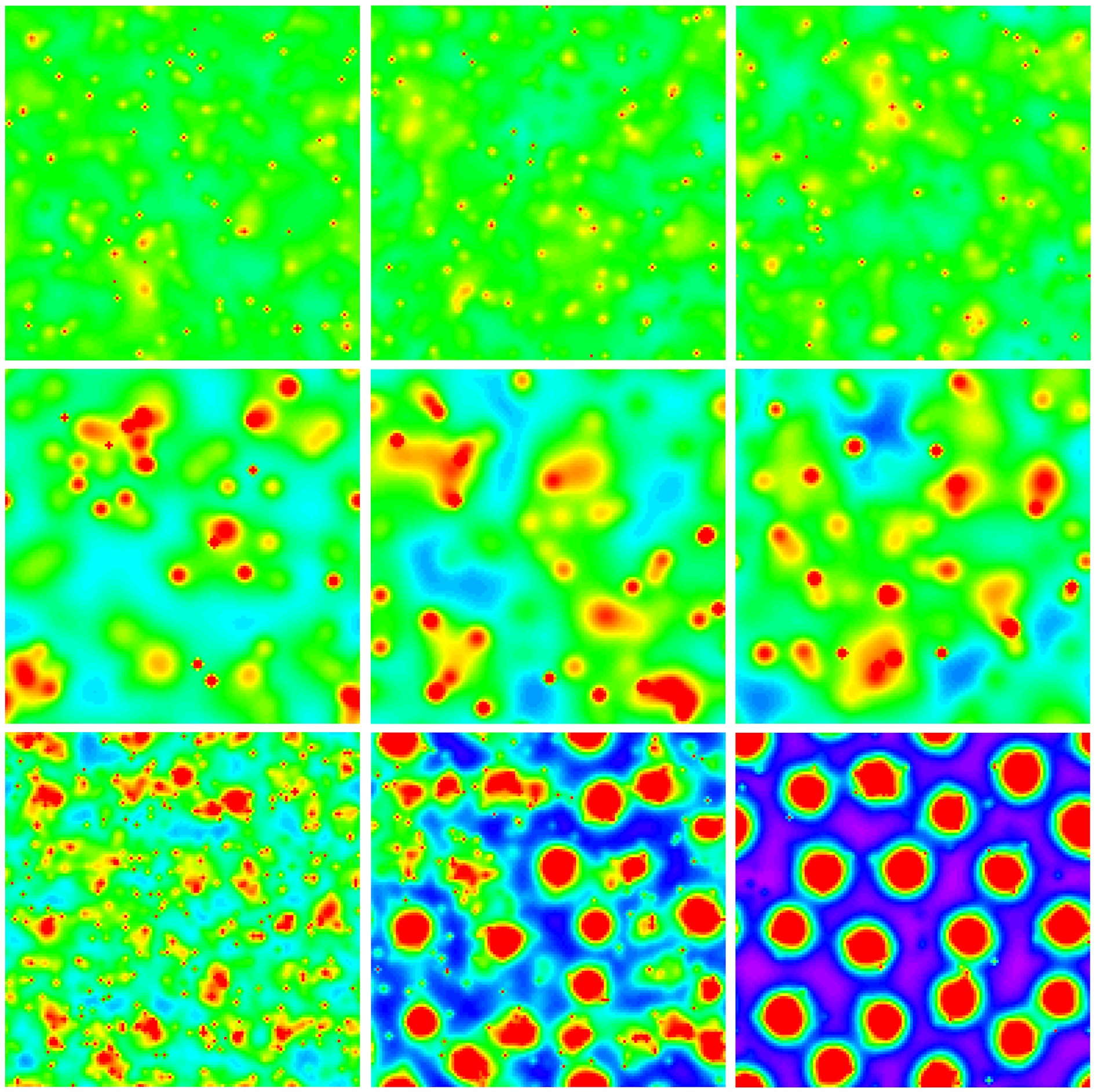}}
\caption{The evolution of crime hotspots. Depending mainly on the number of criminals, hotspots may either not appear at all (top row),  be transient (middle row) or stationary (bottom row). If criminals are few, crime hotspots are likely transient and driven by randomness (middle row), while if criminals are many, the crime hotspots either never appear (top  row) or they do and are stationary (bottom row).  To classify different outcomes, a continuum model is derived and a linear stability analysis is performed, results of which are described in the text. From left to right, the color maps encode the time evolution of the attractiveness field $A_s(t)$ (see Eq.~\ref{field}), such that green represents the midpoint and values below and above follow the rainbow spectrum from violet (minimum) to red (maximum). We refer to \cite{SHORT_MMMAS08}, from where this figure has been adapted, for further details.}
\label{spots}
\end{figure}

Criminal activity is included in the model by allowing individuals to perform one of two actions at every time step. A criminal may either burglarize the site he or she currently occupies, or move to a neighboring one. Burglaries are modeled as random events occurring with probability $p_s(t)=1-\exp[-A_s(t)]$. Whenever site $s$ is burglarized, the corresponding criminal is removed from the lattice, representing the tendency of actual burglars to flee the location of their crime. To balance burglar removal, new criminal agents are generated at a rate $\Gamma$ uniformly on the lattice. If site $s$ is not burglarized, the criminal will move to one of its neighboring sites with probability $1- p_s(t)=\exp[-A_s(t)]$.  The movement is modeled as a biased random walk so that site $s'$ is visited with probability
\begin{equation}
q_{s \to s'}(t)=\frac{A_{s'}(t)}{\sum\limits_{s'}A_{s'}(t)},
\end{equation}
where the sum runs over all neighboring sites of $s$. The position of the criminals and the biasing attractiveness field in Eqs.\,\ref{field} and \,\ref{discreteB} create nonlinear feedback loops which may give rise to complex patterns of aggregation that are reminiscent of actual crime hotspots, similar to those depicted in Fig.~\ref{hot}.
Results for $A_s(t)$ are shown in Fig.~\ref{spots}. Depending on parameter values, it is possible to observe three different behavioral regimes.  In the upper row, all localized increases of $A_s(t)$ that emerge due to recent burglaries disappear very quickly, resulting in a predominantly homogeneous attractiveness field. In the middle row, crime hotspots emerge at random locations and they persist for different periods of time before disappearing or diffusing elsewhere. Lastly, in the bottom row, stationary crime hotspots emerge, which are surrounded by areas of extremely low $A_s(t)$ values. Interestingly, a high number of criminals can result in either the absence of hotspots or intense stationary hotspots. The model actually displays four different regimes of $A_s(t)$ (see Fig.~3 in \cite{SHORT_MMMAS08} for details), but for simplicity we here restrict ourselves to reviewing the three most distinctive cases shown in Fig.~\ref{spots}.

\begin{figure}
\centering{\includegraphics[width = 8.5cm]{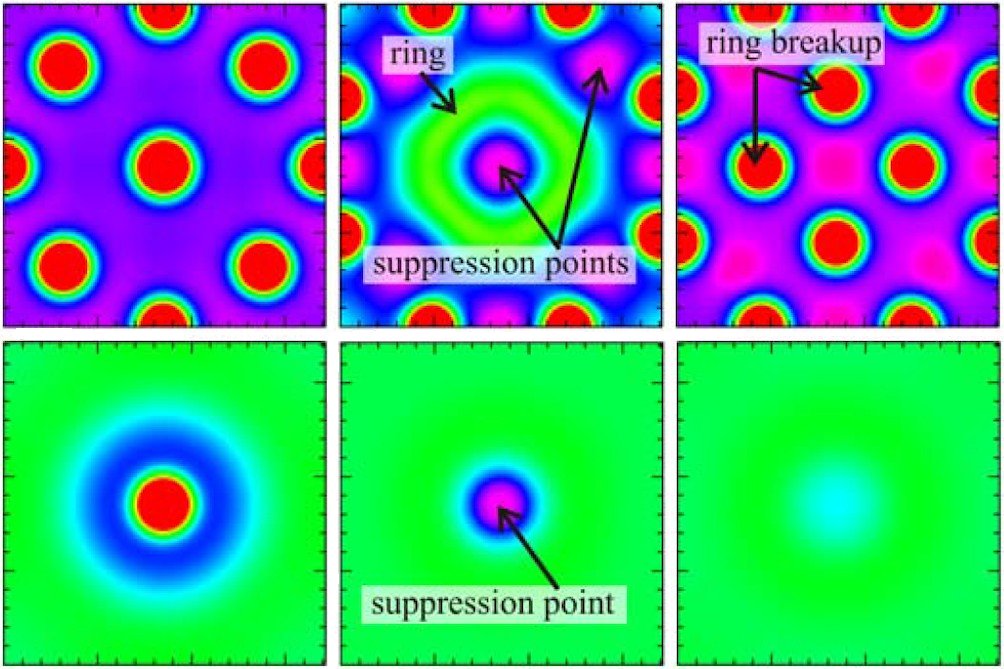}}
\caption{Crime hotspot suppression. Upper row: crime hotspots emerging via a supercritical bifurcation and subjected to suppression. The stationary solution on which the suppression is initiated is shown in the left panel.  The targeted crime hotspots disappear leading to  a transition period characterized by a ``hot ring'' solution around the location of the original central hotspot (middle panel). Eventually,
new hotspots emerge in positions adjacent to the original ones (right panel). Crime has been displaced but not eradicated. Lower row: crime hotspots emerging via a subcritical bifurcation and subjected to suppression. The stationary solution on which the suppression is initiated is shown in the left panel.  The hotspot gradually vanishes without giving rise to new hotspots in nearby locations (middle and right panels).  Crime has been eradicated. All depicted solutions were obtained with Eqs.~\ref{continB} and \ref{continR}. From left to right, the color maps encode the time evolution of the attractiveness $B$, using the same color profile as in Fig.~\ref{spots}. We refer to \cite{SHORT_PNAS10}, from where this figure has been adapted, for further details.}
\label{suppress}
\end{figure}

On the basis of the discrete system it is possible to derive a continuum model, the bifurcation analysis of which yields a more thorough understanding of the spatiotemporal dynamics summarized in Fig.~\ref{spots}.  From the continuum model one can also outline suggestions for crime hotspot suppression and policing \cite{SHORT_PNAS10, SHORT_SIAM10}. The continuum version of the dynamics of the attractiveness field is
\begin{equation}
\frac{\partial B}{\partial t}=\frac{\eta D}{z} \nabla^2B-\omega B + \epsilon D \rho A,
\label{continB}
\end{equation}
where $D=\ell^2/\delta t$, $\epsilon= \delta t$, and $\rho(s,t)=n_s(t)/\ell^2$. Details of the derivation are described in \cite{SHORT_MMMAS08}. The continuum equation for criminal number density, denoted as $\rho$ is given by
\begin{equation}
\frac{\partial \rho}{\partial t}=\frac{D}{z} \vec{\nabla} \cdot \left[\vec{\nabla} \rho - \frac{2 \rho}{A} \vec{\nabla}A  \right] -\rho A +\gamma,
\label{continR}
\end{equation}
where offenders exit the system at a rate $\rho A$, and are reintroduced at a constant rate per unit area $\gamma = \Gamma / \ell^2$. Equations~\ref{continB} and \ref{continR} are coupled partial differential equations that describe the spatiotemporal evolution of the attractiveness $B$ and the offender population $\rho$. They belong to the general class of reaction-diffusion equations that often lead to spatial pattern formation \cite{CROSS_RMP93}.

For a detailed mathematical treatment of Eqs.~\ref{continB} and \ref{continR}, as well as the derivation of their dimensionless form, we refer to \cite{SHORT_MMMAS08, SHORT_PNAS10, SHORT_SIAM10}. Here we summarize the analysis, which shows that parameters used to obtain the first and second row of Fig.\,\ref{spots} correspond to the case where the continuum equations allow for a stable uniform solution, while parameters used to obtain the third row of Fig.\,\ref{spots} correspond to
the case where the uniform solution is unstable. The emerging picture is that crime hotspots form when the enhanced risk of repeat crimes -- measured as a function of all relevant parameters -- is high enough to diffuse locally without binding distant crimes together (for details see Fig.~2 in \cite{SHORT_PNAS10}). Within the unstable regime, the formation of crime hotspots may occur either via supercritical or subcritical bifurcations.  In order to study the effects of police intervention, the crime rate $\rho A$ in Eq.\,\ref{continR} is set to zero at given hotspot locations and for a given time frame \cite{SHORT_PNAS10}.  Numerical studies reveal that only subcritical crime hotspots may be permanently eradicated via the above described suppression mechanism, while supercritical hotspots are only displaced. The two different outcomes are illustrated in Fig.~\ref{suppress}, where the upper and bottom rows show the suppression of supercritical and subcritical crime hotspots, respectively.

Further research on this model include the introduction of spatial disorder,
methods for police suppression to dynamically adapt to evolving crime patterns or to choose from different
deployment strategies and more rigorous analysis \cite{RODRIGUEZ_MMMAS10, JONES_MMMS10, CANTRELL_SIAM12, BERESTYCKI_MMS13,  ZIPKIN_DCDS14}. Other mathematical work on the spread of crime in society include
dynamical systems that include competition between citizens, criminals and guards \cite{NUNO_PHYSA08},
the effects of socio--economic classes, changes in police efficiency and/or resources assigned to them \cite{NUNO_DCDS11}, the effects of imprisonment and recidivism \cite{MCMILLON_PONE14} and the possibility of communities defending themselves from criminals \cite{SOOKNANAN_JMR12}.
Viewed as a whole, this body of work may prove useful in developing better and more cost-effective crime mitigation methods and to allow for the optimization of containment and suppression resources.

\section{Self-exciting point process modeling}
\label{point}
Certain types of crime, like burglary and gang violence, appear clustered in time and space and are reminiscent of earthquake activity. The clustering patterns observed by seismologists indicate that the occurrence of an earthquake is likely to induce a series of aftershocks near the location of the initial event, leading to earthquake swarms and clusters. Similar induction phenomena may be observed in crime pattern formation, motivating the application of seismology methods to model criminal activity. One of these methods is the self-exciting point processes \cite{MOHLER_JASA11}.

A space-time point process is a collection of points representing the location $(x,y)$ and the time $t$ of occurrence of a given event,
such as an earthquake, the striking of lightening or the birth of a species.  The process is associated to a conditional rate $\lambda(x,y,t)$, indicating the rate of occurrence of events at location $(x,y)$ conditioned on the history $H(t)$ of the point process up to time $t$ \cite{DALEY_03}. In seismology, point processes are used by considering a ``parent earthquake" and subsequent  background events or aftershocks. Background activity is modeled as a stationary Poisson process with arrival rate $\mu (x,y)$ that depends on all previous seismic events. Aftershocks are described via a triggering function $g(x,y,t)$ that also depends on previous seismic occurrences, but whose amplitude decreases as a function of the spatio-temporal distance from them. The function $g$ also depends on the magnitude of past earthquakes. These ideas have been translated into criminal modeling by similarly considering ``parent crimes" and subsequent background or offspring crimes. A few modifications in going from earthquake to crime modeling are necessary such as the introduction of a multiplicative factor $\nu(t)$ in the background activity which embodies global fluctuations
due to weather, seasonality or time of day. Also, while in seismology decades of research and refinement have lead to well defined functional forms for $g$, in crime non-parametric methods and calibrations using actual data are used to estimate $g$ as well  $\nu$ and $ \mu$. For details on the forms of $\mu$ and $g$ used in seismology and on the iterative procedures used in crime we refer to the seminal work by Mohler et al.~\cite{MOHLER_JASA11}.

The application of self-exciting point process modeling to urban crime has been tested using residential burglary data provided by the Los Angeles Police Department ~\cite{MOHLER_JASA11}. Previous methods of predicting crime had been introduced in the literature \cite{BOWERS_04} where crime hotspot maps were generated via a pre-assigned fixed kernel, using previous crime occurrences as input. The point process methodology has been found superior to the crime hotspot map, even for robberies or auto theft where near-repeat effects play less of a role. The main advantage of point process modeling has been attributed to a better balance between exogenous and endogenous contributions to crime rates and to its direct inference from data rather than an a priori imposition of hotspot maps. Accordingly, the usage of self-exciting point process modeling is in many ways superior to using a pre-assigned fixed kernel.

Self-exciting point processes have also been used to analyze temporal patterns of civilian death reports in Iraq between 2003 and 2007 \cite{LEWIS_SJ12}. Similarly to urban crimes, the rate of violent events has been partitioned into the sum of a Poisson background rate and a self-exciting component in which previous bombings or other episodes of violence generate a sequence of offspring events according to a Poisson distribution. Comparing with actual data, the study showed that point processes are well suited for modeling the temporal dynamics of violence in Iraq. Point processes may also be used in geographic profiling of criminal offenders to estimate the probability density for the home base of a criminal who has committed a given set of spatially distributed crimes. Target selection from a hypothetical home base is informed by geographic inhomogeneities such as housing types, parks, freeways
or other physical barriers as well as directional bias and preferred distances to crime \cite{MOHLER_SIAM12}. In the context of geographic profiling, point processes are used to estimate the crime probability density given a set of spatially distributed crimes.
In more recent work, self-exciting point processes have been used to model intra-gang violence due to retaliation after an initial
attack \cite{SHORT_DCDS14}.

Overall, the application of self-exciting point processes, inspired by earthquake prediction in seismology, can be used successfully for modeling and predicting crime, as demonstrated on residential burglary data from the Los Angeles Police Department~\cite{MOHLER_JASA11} and on temporal patterns of civilian death reports in Iraq~\cite{LEWIS_SJ12}. In the future it may be desirable to further tailor point process models specifically for crime type and local geography and to refine the construction of parametric models which could facilitate the application of this promising methodology.

\section{Crime as a social dilemma}
\label{social}
Evolutionary game theory \cite{MAYNARD_82, HOFBAUER_98, MESTERTONG_01, NOWAK_06, SIGMUND_10} has been the traditional framework of choice for studying the evolution of different behavioral strategies in a competitive setting \cite{SANTOS_N08, SANTOS_PNAS11, FU_SREP12, FU_SREP12B}. From the large array of possible games, few have received as much attention as the prisoner's dilemma \cite{AXELROD_84}. Here, within each round, two players must decide simultaneously whether they want to cooperate with each other or not. Each player then receives a payoff that depends on the mutual decisions made. A ``social dilemma'' arises because cooperation between both players would yield the highest collective payoff, but the payoff for a defector is higher if the opponent decides to cooperate. Mutual defection is therefore the only rational outcome that emerges if both players act selfishly so as to maximize their individual profit. In the long run, the proliferation of defection inevitably leads to the ``tragedy of the commons'' \cite{HARDIN_G_S68}, where common resources are lost to society due to overexploitation and lack of cooperative care.

\begin{figure}
\centering{\includegraphics[width = 5cm]{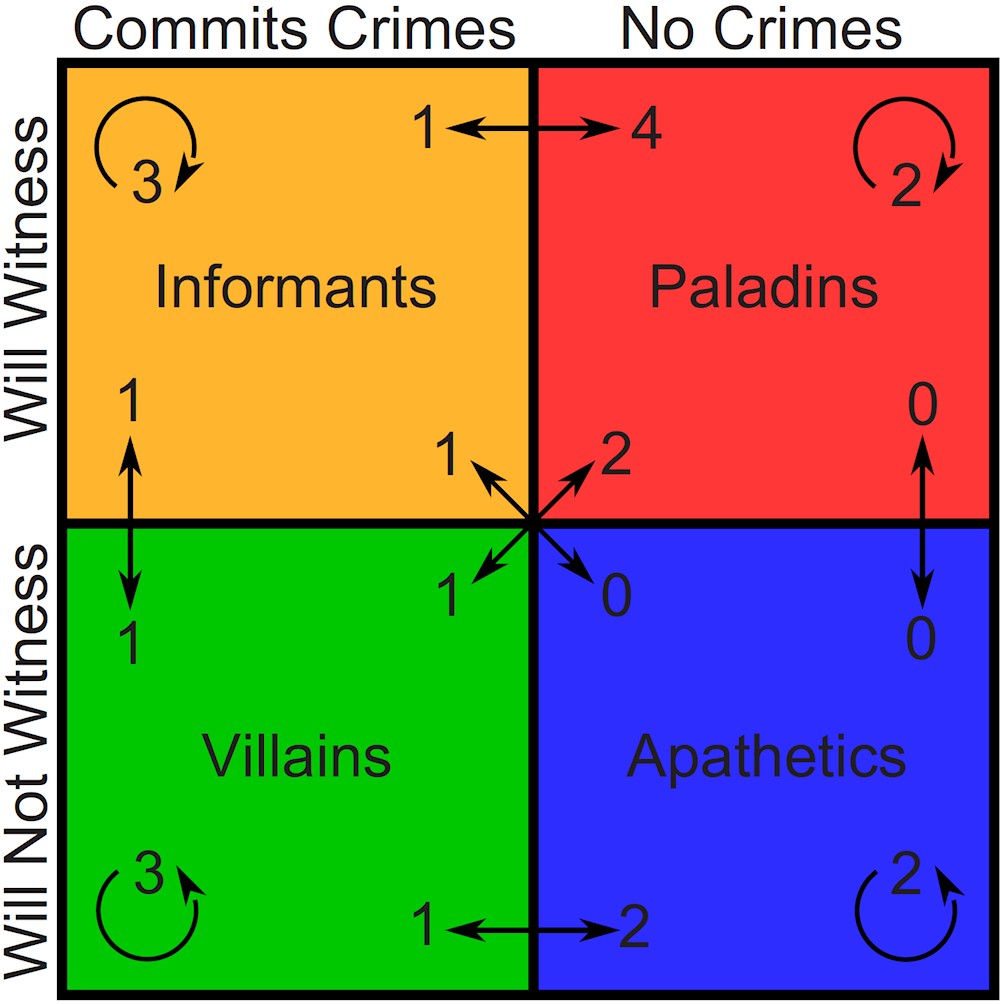}}
\caption{Crime as an evolutionary game. The society is composed of four strategies -- informants, paladins, villains and apathetics -- defined by their propensities to both commit crimes and serve as witnesses in criminal investigations. Arrows between strategies indicate the number of possible game pairings and outcomes in which the update step leads to a strategy change. For example, there are two ways by means of which a villain can be converted into a paladin. Circular arrows within each strategy quadrant indicate updates
where player strategies remain unchanged. This figure has been reproduced from \cite{SHORT_PRE10}.}
\label{strats}
\end{figure}

Although criminal behavior does not necessarily map to the prisoner's dilemma, it is nevertheless possible, and indeed very rewarding, to study the evolution of crime within the framework of social dilemmas \cite{NOWAK_S06}. In this context, social order can be considered as the common good that is threatened by criminal activity, with competition arising between criminals and those trying to prevent crime. However, committing crimes is not necessarily equivalent to defection, since criminals, unlike defectors, may actively seek to harm others. By the same token, fighting crime can be more than simply to cooperate, in particular since it may involve risk that goes beyond contributing some fraction of one's ``wealth'' into the common pool. Although in principle committing crime and defecting, as well as fighting crime and cooperating are in good correspondence, a more deliberate formulation of the competing strategies may elevate the accuracy of the modeling approach.

With these considerations in mind, an adversarial evolutionary game including four competing strategies can be constructed \cite{SHORT_PRE10} as summarized in Fig.~\ref{strats}. The game entails informants ($I$) and villains ($V$) as those who commit crimes, as well as paladins ($P$) and apathetics ($A$) as those who do not. Informants and paladins actively contribute to crime abatement by collaborating with authorities whenever called upon to do so. All players may witness crimes or be the victims of crime, in agreement with victimization surveys \cite{LYNCH_07}. Thus, paladins are model citizens that do not commit crimes and collaborate with authorities, while villains, to the other extreme of the spectrum,  commit crimes and do not report them. Intermediate figures are informants who report on other offenders while still committing crimes, and apathetics who neither commit crimes nor report to authorities. The lack of active cooperation in apathetics may be due to inherent apathy, fear of retaliation or ostracism from the community at large. Apathetics are similar to second-order free-riders in the context of the public goods game with punishment \cite{FEHR_N04, SIGMUND_TREE07}, in that they cooperate at first order by not committing crimes, but defect at second order by not punishing offenders.

The game unfolds iteratively. At each round a criminal is randomly selected from the $V+I$ pool together with a potential victim from the $N-1$ remainder of the population. The two selected players begin the game with a unitary payoff. After a crime occurs, the criminal player increases its payoff by $\delta$, while the victim looses $\delta$. If the victim is either an apathetic or a villain, the crime is not reported to the authorities and therefore successful: the victim's payoff is decreased to $1-\delta$ and the victimizer's is increased to $1+\delta$. If, on the other hand, the victim is a paladin or an informant, the crime is reported to the authorities and an ``investigation'' begins. For this, a subset $M$ of the $N-2$ remaining players is drawn, and the victimizer is convicted with probability $w=(m_P+m_I)/M$, where $m_P$ and $m_I$ are the number of paladins and informants within $M$. In case of a conviction, the victim is refunded $\delta$, and payoff of the criminal becomes $1-\theta$, where $\theta$ determines the severity of punishment. With probability $1-w$ the crime is left unpunished, in which case the criminal retains $1+\delta$, while the victim's payoff is further decreased to $1-\delta-\epsilon$. Here $\epsilon$ may be interpreted as retaliation on the accuser as perpetrated by the accused who, having escaped punishment, feels empowered in his or her revenge. Other interpretations of $\epsilon$ may be damages to personal image or credibility, or a loss of ``faith in the system'' after making an accusation that is unsubstantiated by the community. Note that in the latter case, the choice of reporting one's victimization to authorities may be even more detrimental to the witness than the original criminal act ($\epsilon>\delta$).  This scenario especially applies to societies heavily marred by war, by mafia or drug cartels, where few people will serve as witnesses to crimes. Parameters $\delta$, $\theta$ and $\epsilon$ are always used such that all payoffs remain positive. At the end of each round of the game, the player with the smaller payoff changes his or her strategy according to proportional imitation \cite{SCHLAG_JET98}. In particular, if the victimizer is emulated, the loser simply adopts the victimizer's strategy and ends the update as either a villain or an informant. If the victim is emulated, the loser mimics the victim's propensity to serve as a witness but adopts a noncriminal strategy regardless of the victim's. In this case, the update results with the loser becoming either a paladin or an apathetic (see Fig.~\ref{strats} for details).

\begin{figure}
\centering{\includegraphics[width = 6.4cm]{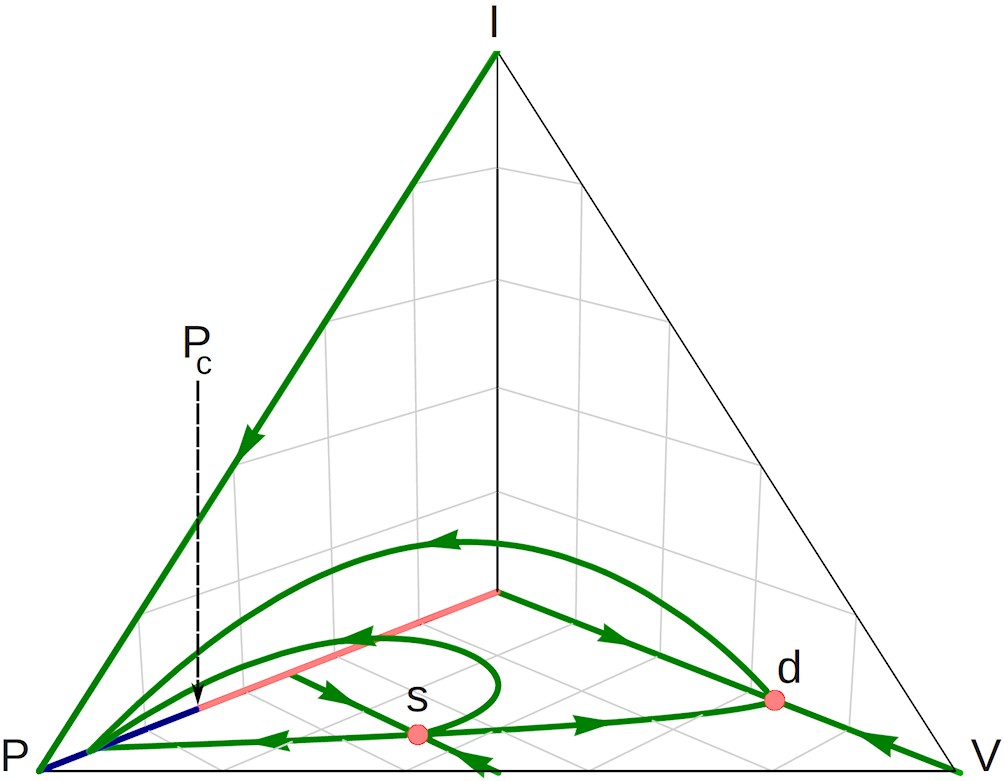}}
\caption{The emergence of utopia in a society with informants. All trajectories with $I_0>0$ evolve towards a crime-free state. The phase diagram shows unstable fixed points in light red, unstable fixed lines in thick light red, stable fixed lines in thick dark blue, and trajectories beginning (or ending) along various eigenvectors as thick green arrows. The dystopian fixed point $d$ and the saddle point $s$ are unstable to increases in $I$, so that the only attracting final states for $I_0>0$ are those utopias with $P>P_c$. These results were obtained with $\delta=0.3$, $\theta=0.6$ and $\epsilon=0.2$, but qualitative results are independent of parameters. This figure has been reproduced from \cite{SHORT_PRE10}.}
\label{ternary}
\end{figure}

Stochastic simulations reveal that informants are key to the emergence of a crime-free society -- ``utopia". Indeed, a crime-dominated society can be transitioned to one that is largely crime-free by imposing an optimal number of informants $I_0$ at the onset of the game. The dynamics depend on the chosen parameterizations and utopia may be more difficult to achieve in an extremely adversarial society, with initial high numbers of villains and apathetics. However, by deriving a deterministic version of the above described game \cite{SHORT_PRE10} it is possible to show that if there are at least some informants initially present in the population ($I_0 > 0$)
the final state is always utopia regardless of $\delta$, $\theta$ and $\epsilon$. This is illustrated in Fig.~\ref{ternary}, which features a 4D ternary phase diagram of the four competing strategies.

While beneficial, the presence of informants may come at a cost, either in training an undercover informant, or in convincing a criminal to collaborate with authorities, or in tolerating the criminal acts that informants will keep committing. One may thus consider an optimal control problem \cite{SHORT_EJAM13} to investigate the active recruitment of informants from the general population in terms of associated costs and benefits. Higher recruitment levels may be the most beneficial in abating crime, but they may also be prohibitively expensive.  Recruitment costs are designed to depend on the past history of players so that the conversion of individuals with higher cumulative past payoffs might be more costly than that of less successful ones \cite{SHORT_EJAM13}.  The optimal control problem was expressed via three control functions subject to a system of delay differential equations, and was solved and discussed under different settings. Targeted and random recruitment of informants were also considered. Optimal recruitment strategies were shown to change drastically as parameters and resource constraints were varied and that more information about individual player strategies leads only to marginally lower costs.

\begin{figure}
\centering{\includegraphics[width = 8.5cm]{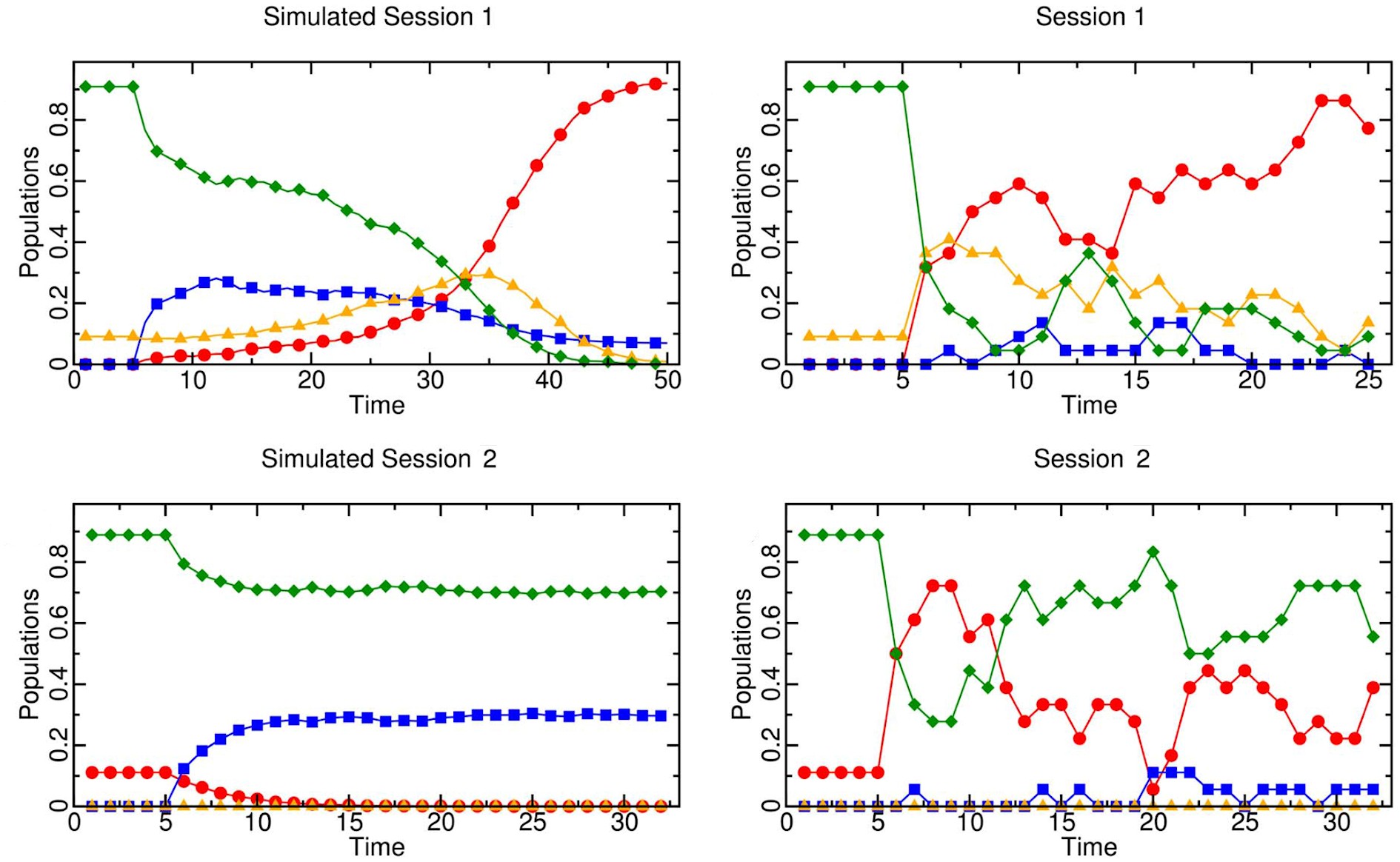}}
\caption{Human experiments confirm that informants are key to diminishing crime. Depicted are comparisons of the strategy evolutions simulated from theory (left) and obtained from experimental sessions (right). In the top row all four strategies are allowed, while in the bottom row informants are disallowed. Both simulation and human experiment outcomes confirm that the absence of informants leads to elevated levels of criminal behavior. Future challenges remain to determine how well the model and the human experiment actually fit to a potential real-life scenario. Paladins are red circles, apathetics are blue squares, informants are orange triangles, and villains are green diamonds. For further details we refer to \cite{DORSOGNA_PONE13}, from which this figure was adapted.}
\label{experiment}
\end{figure}

The crucial role of informants within the reviewed adversarial evolutionary game \cite{SHORT_PRE10} has also been studied by means of human experiments in \cite{DORSOGNA_PONE13}. The goal was to test whether informants are indeed critical towards crime abatement as predicted by theory. Quite remarkably, as illustrated in Fig.~\ref{experiment}, good agreement between the outcome of the stochastic simulations and the laboratory experiments was obtained for different parameterizations of the game. Human experiments thus confirmed that reaching and maintaining a low-crime society may be favored by seeking cooperation with active criminals. Details on adaptations of the theoretical game to a laboratory setting and nuanced considerations on the role of informants in the two settings are described in  \cite{DORSOGNA_PONE13}.

The evolution of crime can also be studied through the lens of the inspection game \cite{TSEBELIS_RS90}. Rational choice theories predict that increasing fines should diminish crime \cite{BECKER_JPE68}. However, a three strategy inspection game, where in addition to criminals ($C$) and punishing inspectors ($P$), ordinary individuals ($O$) are present as well leads to very different outcomes \cite{PERC_PONE13} than what expected. The $O$ players neither commit crimes nor participate in inspection activities and represent the ``masses'' that catalyze rewards for criminals and costs for inspectors. Ordinary individuals receive no bonus payoffs upon encountering inspectors or their peers. Only when paired with criminals do they suffer the consequences of crime in form of a negative payoff $-g \le0$. Criminals, vice-versa, gain the reward $g\ge 0$ for committing a crime. When paired with inspectors criminals receive a payoff $g-f$, where $f\ge 0$ is a punishment fine. When two criminals are paired none of the two are assigned any benefits. Inspectors, on the other hand, always have the cost of inspection, $c \ge 0$, but when confronted with a criminal, an inspector receives the reward $r\ge 0$ for a successful apprehension. This game was studied via Monte Carlo simulations on a square lattice with periodic boundary conditions where each lattice site is occupied either by a criminal, a punishing inspector, or an ordinary citizen. The game evolves by first randomly selecting player $s$ to play the inspection game with its four nearest neighbors, yielding the payoff $P_{s}$. One of the nearest neighbors of player $s$, $s'$ is now chosen randomly to play the game with its nearest neighbors, leading to $P_{s'}$ analogously to player $s$ before. Finally, player $s'$ imitates the strategy of player $s$ with probability
\begin{equation}
q=\frac{1}{1+\exp(P_{s'}-P_{s})/K]},
\label{MNL}
\end{equation}
where $K$ determines the level of uncertainty in the strategy adoption process. The chosen form in Eq.~\ref{MNL} corresponds to the empirically supported multinomial logit model \cite{MCFADDEN_74}, which for two decision alternatives is also known as the Fermi law \cite{SZABO_PR07, SZOLNOKI_PRE09}. A finite value of $K$ accounts for the fact that better performing players are readily imitated, although it is not impossible to adopt a player performing worse, for example due to imperfect information or errors in decision making.

\begin{figure}
\centering{\includegraphics[width = 8.5cm]{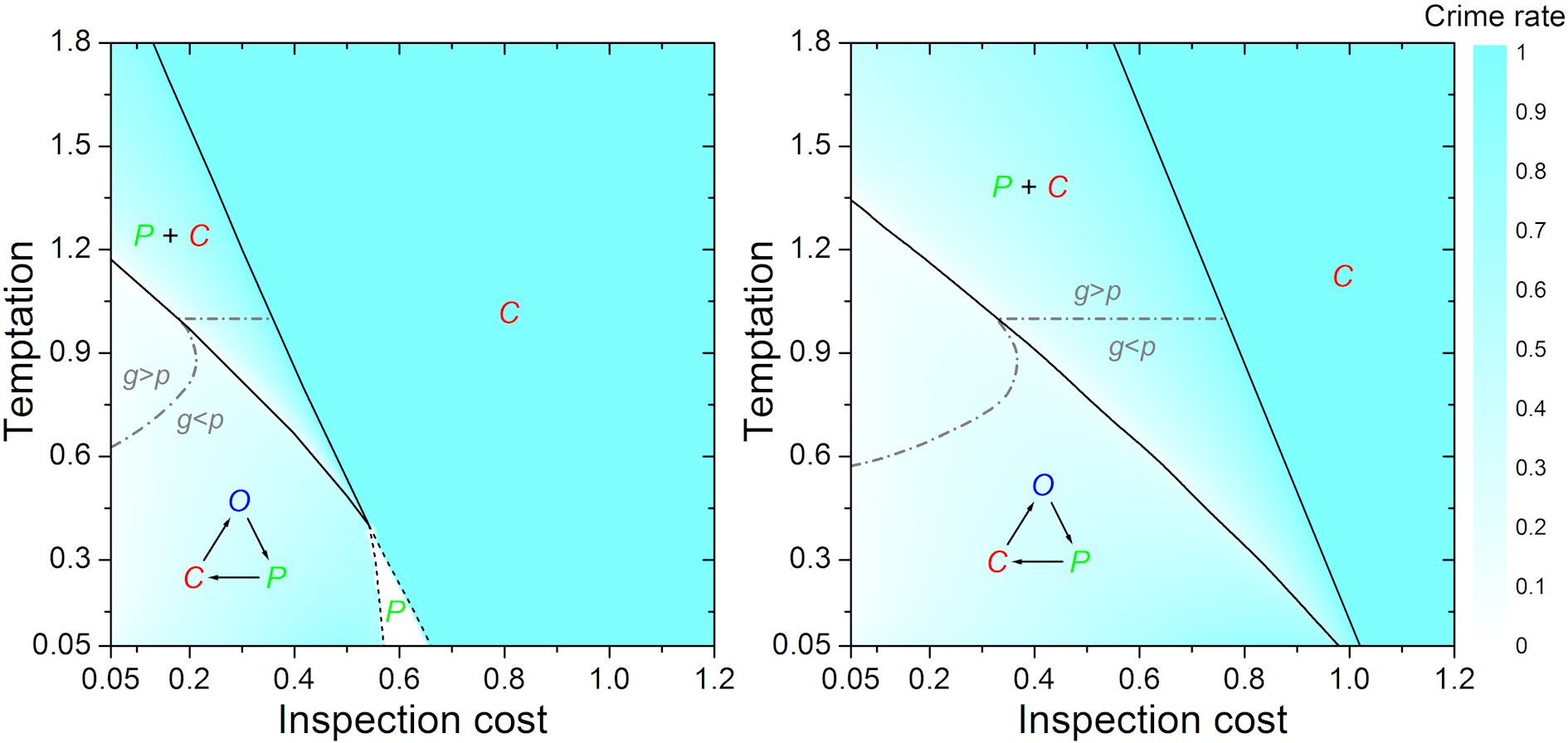}}
\caption{Phase diagrams demonstrating the spontaneous emergence and stability of the recurrent nature of crime and other possible outcomes of the evolutionary competition between criminals ($C$), ordinary people ($O$) and the police ($P$). The diagrams show the strategies remaining on the square lattice after sufficiently long relaxation times as a function of the inspection cost $c$ and the temptation to commit crime $g$, for low (left) and high (right) reward of successful inspection $r$. The overlayed color map encodes the stationary density of criminals in the population (crime rate). For small and intermediate values of $c$ and $g$, cyclic dominance between the three strategies characterizes the evolutionary dynamics. Criminals outperform ordinary people, ordinary people outperform the police, and the police outperform criminals. This cyclic dominance leads to recurrent outbreaks of crime during the evolutionary process. If either $c$ or $g$ exceed a certain threshold, the cyclic phase ends with a continuous phase transition to a mixed $P+C$ phase (lower solid line), where police and criminals coexist. Further increasing the two parameters leads to another continuous transition (upper solid line) and an absorbing $C$ phase, where criminals dominate. A re-entry into the cyclic $C+O+P$ phase is possible through a succession of two discontinuous phase transitions (dashed lines) occurring for sufficiently small $g$ and decreasing inspection costs. First, the absorbing $C$ phase changes abruptly to an absorbing $P$ phase dominated by inspectors, which then changes abruptly to the cyclic phase. If $r$ is larger (right), the region of cyclic dominance increases, but the possibility of complete dominance of the police also vanishes. Dash-dotted gray lines correspond to the condition where the probability for criminals to be detected $p$ is the same as the temptation to commit crime, and a transition to criminal behavior would thus be expected according to the rational choice theory. For further details we refer to \cite{PERC_PONE13}, from where this figure was adapted.}
\label{phase}
\end{figure}

Monte Carlo simulations reveal that the collective behavior of the three-strategy spatial inspection game is complex and counterintuitive. As depicted in Fig.~\ref{phase}, continuous (solid lines) and discontinuous (dashed lines) transitions between different phases emerge. For the left panel of Fig.~\ref{phase}, where we use low reward values $r$ for successful inspection, these include (i) a dominance of criminals for high rewards of committing a crime $g$ and high inspection costs $c$ (the $C$ phase), (ii) a coexistence of criminals and punishing inspectors for large values of $g$ and moderate values of $c$ (the $P+C$ phase), (iii) a dominance of punishing inspectors for moderate inspection costs and low values of $g$ (the $P$ phase), and (iv) cyclical dominance for small inspection costs and small values of $g$ (the $C+O+P$ phase). In the cyclic dominance phase criminals beat ordinary individuals, ordinary individuals beat punishing inspectors, and punishing inspectors win against the criminals. Noteworthy, the $C+O+P$ phase yields oscillatory behavior, where $C$ beat $O$, $O$ beat $P$, and $P$ beat $C$, thus closing the dominance loop ($C \to O \to P \to C$). The cyclic dominance that is responsible for strategy density oscillations emerges spontaneously due to pattern formation and is robust against initial condition variations.

The three-strategy inspection game shows that the interplay between criminal activities and sanctioning efforts may be complex and lead
to non-linear dynamics that may make it difficult to devise successful intuitive crime prevention policies. Other game-theoretic optimization studies include finding the best way to defend multiple sites against criminal attempts who also must choose among several attack locations \cite{PRIMICERIO_PHYSA13}. Results from this section indicate that crime should be viewed not only as the result of offending actions committed by certain individuals, but also as the result of social interactions between people who adjust their behavior in response to societal cues and imitative interactions. The emergence of crime thus should not be ascribed merely to the ``criminal nature'' of particular individuals, but rather to the social context, the systems of rewards and punishment, the level of engagement of the community, as well as to the interactions between individuals. This more comprehensive view of crime may have relevant implications for policies and law enforcement.

\section{Networks of crime, gangs and geography}
\label{gangs}
A natural extension of the research reviewed thus far, where the focus has been on crime hotspots and the propensity of single individuals to commit crime, is to consider criminal networks, the formation of gangs, and the geography of crime building, and in particular how these topics could benefit from recent advances in network science \cite{ALBERT_RMP01, DOROGOVTSEV_AP02, BOCCALETTI_PR06, havlin_pst12, boccaletti_14, kivela_jcn14}. Forms of large-scale organized crime \cite{MALLORY_11}, such as the Italian Mafia \cite{GAMBETTA_88}, street gangs, or drug cartels \cite{BEITTEL_09} usually emerge when fear and despair become so ingrained within a society that the social norm is simply to accept crime, so that witnesses and even victims of crime choose not to cooperate with law enforcement in the prosecution of criminals. Instead, one tries to fit in. Acquiescence and acceptance are usually slippery slopes towards later forms of active engagement, leading to the growth of the local criminal organization or a criminal network. Criminological research has identified a number of factors that may promote the regional development of crime, including unemployment \cite{RAPHAEL_JLE01, LIN_JHR08}, economic deprivation \cite{LAFREE_ARS99}, untoward youth culture \cite{CURTIS_JCLC97}, failing social institutions \cite{LAFREE_98}, issues with political legitimacy \cite{LAFREE_ARS99}, as well as lenient local law enforcement strategies \cite{CORSARO_JEC09, MCGARRELL_JQC10}, to name but a few examples. Recent work on declining criminal behavior in the U.S. in fact suggests that trends in the levels of crime may be best understood as arising from a complex interplay of many such factors \cite{GOMEZ_JAE06, ZIMRING_06}, while most recent empirical data indicate that social networks of criminals have a particulary strong impact on the occurrence of crime -- the more the criminals are connected into networks, the higher the crime rate \cite{PAPACHRISTOS_JUH12, PAPACHRISTOS_ASR13}.

\begin{figure}
\centering{\includegraphics[width = 7.9cm]{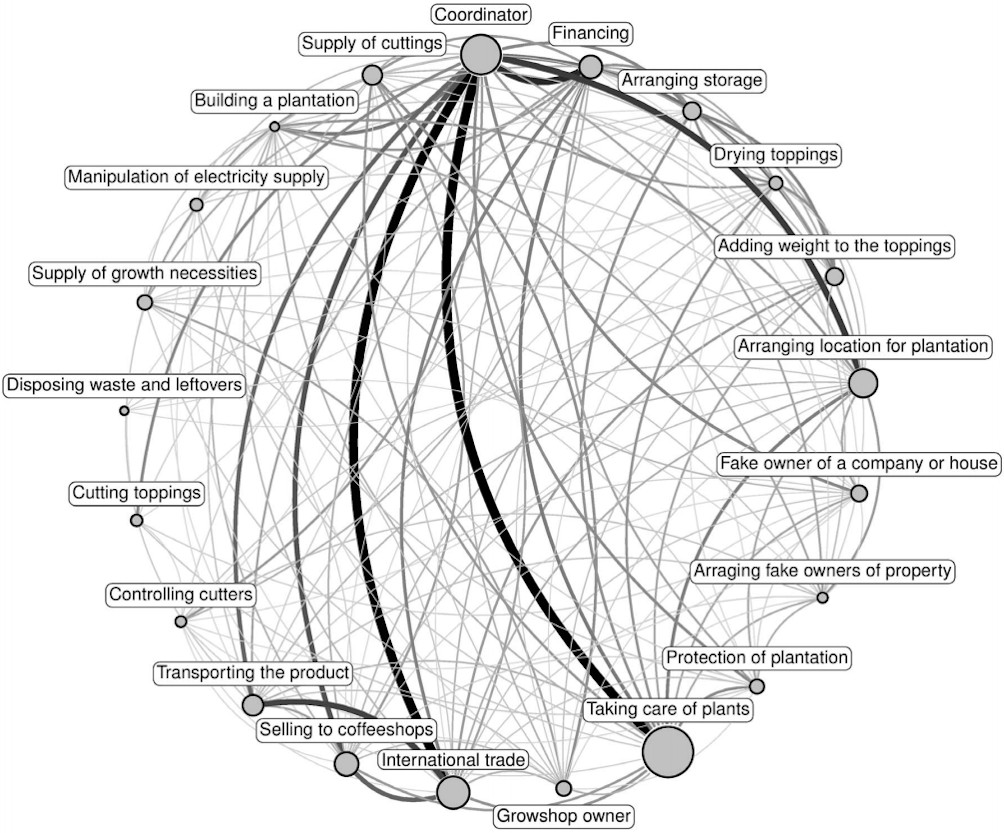}}
\caption{A cannabis cultivation criminal network in the Netherlands.  Nodes represent the many actors needed for successful production and distribution of cannabis. The network is highly resilient to targeted disruption strategies and perturbations will lead to reorganization towards a more robust and resilient network. Node sizes represent the number of actors fulfilling the associated role, and link thickness corresponds to the total number of links between actor groups. For further details we refer to \cite{DUIJN_SR14}, from where this figure has been reproduced.}
\label{cannabis}
\end{figure}

The assumption that there is a network structure behind organized crime immediately invites the idea that removing the leader, or the most important hubs of the network \cite{ALBERT_N00}, will disrupt the organization to its very core and thus hinder further criminal activity. Police thus often attempt to identify and arrest the ``ring leader'' of the targeted criminal organization. But even if successful, such operations rarely have the desired effect. A recent study analyzing cannabis production and distribution networks in the Netherlands shows that this strategy may be fundamentally flawed \cite{DUIJN_SR14}. As depicted in Fig.~\ref{cannabis}, all attempts towards network disruption analyzed in the study proved to be at best unsuccessful, at worse increased the efficiency of the network, which through nifty reorganizations and recovery ultimately became stronger. By combining computational modeling and social network analysis with unique criminal network intelligence data from the Dutch Police, Duijn et al.~\cite{DUIJN_SR14} have concluded that criminal network interventions are likely to be effective only if applied at the very early stages of network growth, before the network gets a chance to organize, or to reorganize to maximum resilience.

Gangs are similar to criminal networks, although their activity is usually more geographically constrained and segregated \cite{SCHELLING_78, GOLDSTEIN_ORE03}, and their organization features less hierarchy and complexity. The seminal work by Schelling on dynamic models of segregation \cite{SCHELLING_JMS71} and subsequent variations \cite{ZHANG_JMS04, FOSSETT_JMS06, MACY_JMS06} considered agent-based modeling on a square lattice to take into account structured time-invariable interactions \cite{PERC_JRSI13}. The consideration of a structured rather than a well-mixed population is crucial because, in a criminal network or a gang, not everybody is connected to everybody else, and the interactions among members usually follow an established pattern that does not vary over time. Although the usage of realistic social networks might be even more appropriate, the square lattice is a good first-order approximation. The latter allow for the implementation of statistical physics methods \cite{BINDER_88} which have long been used to analyze related systems of interacting particles \cite{LIGGETT_85}.

\begin{figure}
\centering{\includegraphics[width = 8.5cm]{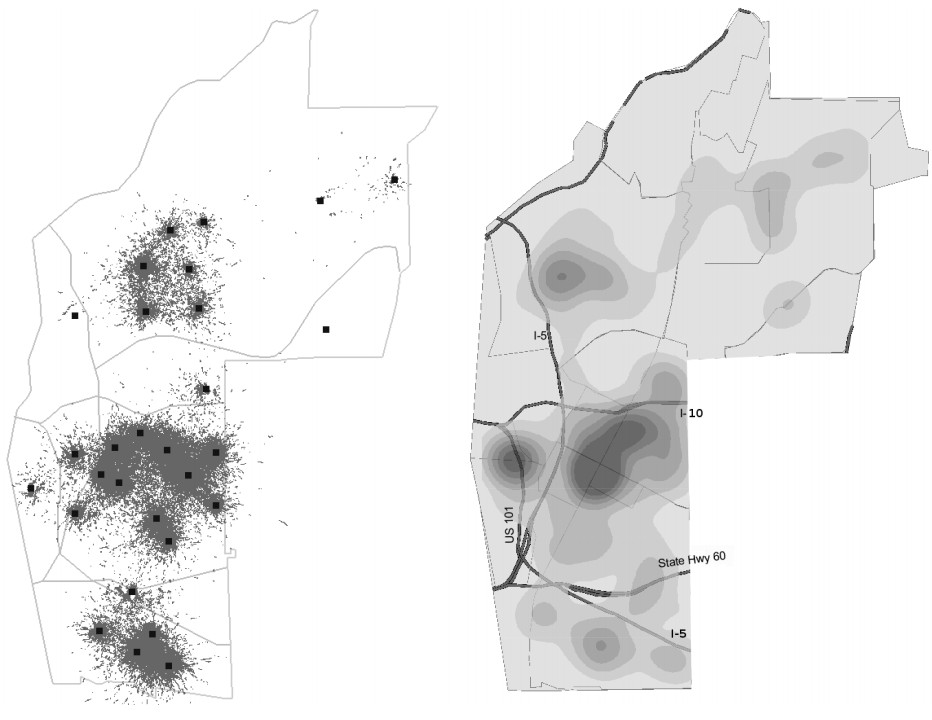}}
\caption{Reconstructing a gang network. Interactions between agents simulated using the biased L\'evy walk network method  (left) . Actual density map of gang-related violent crimes in Hollenbeck between $1998$ and $2000$ (right). Thick lines represent major freeways crossing the city. Further details are described in \cite{HEGEMANN_PA11}, from where this figure has been reproduced.}
\label{SBLN}
\end{figure}

The creation of street gang rivalries was studied via agent-based simulations in conjunction with
data from the Hollenbeck policing division of the Los Angeles Police Department \cite{HEGEMANN_PA11}, home to many urban gangs.
Each agent is part of an evolving rivalry network that includes past interactions
between gang members. Individuals perform random walks where the jump length is drawn from a truncated L\'evy distribution and where bias in the direction of rivals is included. Gang home bases, historical turfs and geographic details that may limit movement such as freeways, rivers and parks are included in the so called simulated biased L\'evy walk network, as well as typical gang behaviors inferred from the criminology literature. Simplified baseline models are also simulated and results from all models are compared with actual gang networks in Hollenbeck.
Using metrics derived from graph theory, it is possible to show that simulated biased L\'evy walk network modeling is the most accurate in replicating the actual gang network, compared to the other, more simple methods. For comparison we show simulated results and an actual map of violent crimes in Hollenbeck in Fig.~\ref{SBLN}. Furthermore, the simulated biased L\'evy walk network converges to stable long-term configurations, which is useful when modeling unknown rivalry interactions. The method is portable and can be applied to other geographical locations, offering insight on gang rivalry distributions in the absence of known data. It may also be extended to test sociological concepts related to gang interactions such as territoriality and/or allegiances within gangs.

Police department field interview cards were later used to study the behavioral patterns of roughly $748$
suspected gang members who were stopped and questioned in Hollenbeck \cite{VanGennip_SIAM13}. The goal was to
identify any social communities among street gang members by creating a fully--connected ad hoc graph
where individuals represent nodes and links encode geographical and social data.
Individuals stopped together were assumed to share a friendly or social link and the distance $d_{i,j}$ between
stop locations of individuals was recorded.
This information was used to determine the affinity matrix $W_{i,j}$ associated with the graph. Its entries are composed of a term that decays  as a function of $d_{i,j}$, representing geographical information,
and of an adjacency matrix whose entries are zero or one depending on whether individuals
were stopped together or not. The latter represents social information.
Using spectral clustering methods distinct groups were identified and interpreted
as distinct social communities among Hollenbeck gang members. These
clustered communities were then matched with actual gang affiliations recorded from the
police field interview cards.  To evaluate cluster quality the authors use a purity measure, defined as  the number of correctly identified gang members in each cluster divided by the total number of gang members.
Results showed that using geographical information alone leads to
clustering purity of about 56$\%$ with respect to the true affiliations of the 748  individuals taken in consideration.
Adding social data may improve purity levels, especially if this data is used
in conjunction with other information, such as friendship or rivalry networks. These results may be used
as a practical tool for law enforcement in providing useful starting points when trying to identify possible culprits
of a gang attack.

A mathematical approach to modeling gang aggregation and territory formation by means of an Ising-like model on a square lattice has recently also been proposed in \cite{BARBARO_PA13}. Here, otherwise indistinguishable agents are allowed to aggregate within two distinct gangs and to lay graffiti on the sites they occupy. Interactions among individuals are indirect and occur only via the graffiti markings present on-site and on nearest-neighbor sites.  Graffiti is subject to decay either from the elements or from active police removal. Within this model, gang clustering and territory formation may arise under specific parameter choices, and a phase transition may occur between well-mixed, possibly dilute configurations and well separated, clustered configurations. Using methods of statistical physics, the phase transition between these two qualitatively different scenarios has been studied in detail. In the mean-field rendition of the model, parameter regimes were identified where the transition is first or second order. In all cases however, these clustering transitions were driven by gang-to-graffiti couplings since direct gang-to-gang interactions were not included in the model. This leads to the conclusion that indirect coupling between gangs, such as graffiti markings, may be the sole catalyst for gang clustering. The role of graffiti and vandalism has been recently reviewed by Thompson et al.~\cite{THOMPSON_A12}, who analyzed the urban rail industry, where graffiti markings have significant impact on expenditure, timely operation of services, and on passenger perception of safety.

Lastly, we mention promising efforts to detect criminal organizations \cite{FERRARA_ESA14} and to predict crime \cite{bogomolov_acm14} based on demographics and mobile data. It is known that the usage of communication media such as mobile phones and online social networks leaves digital traces, and research shows that this data can be used successfully for detecting and characterizing criminal organizations. With the aid of statistical network analysis and community detection \cite{fortunato_pr10}, recent advances could allow forensic investigators to better understand hierarchies within criminal organizations, to discover members who play central role, as well as provide valuable information on connections among sub-groups \cite{FERRARA_ESA14}.

\section{Rehabilitation and recidivism}
\label{rehab}
Perhaps most fitting to end this review, we turn to rehabilitation and recidivism as successful and unsuccessful ends to the treatment of past offenders, respectively. In addition to punishing wrongdoers, the justice system should aim to rehabilitate and integrate past offenders into society. Recidivism is a sign that this process has failed, as offenders who experience punishment relapse into crime. Here, the dilemma is that of the ``stick versus carrot'', a dilemma that has already received ample attention within evolutionary public goods game \cite{SIGMUND_PNAS01}, the main focus being on punishment \cite{SIGMUND_TREE07}. On the other hand, recent research on antisocial punishment has raised concerns on the use of sanctions as a means to promote collaborative efforts and to raise social welfare \cite{HERRMANN_S08, RAND_S09}.

\begin{figure}
\centering{\includegraphics[width = 8.5cm]{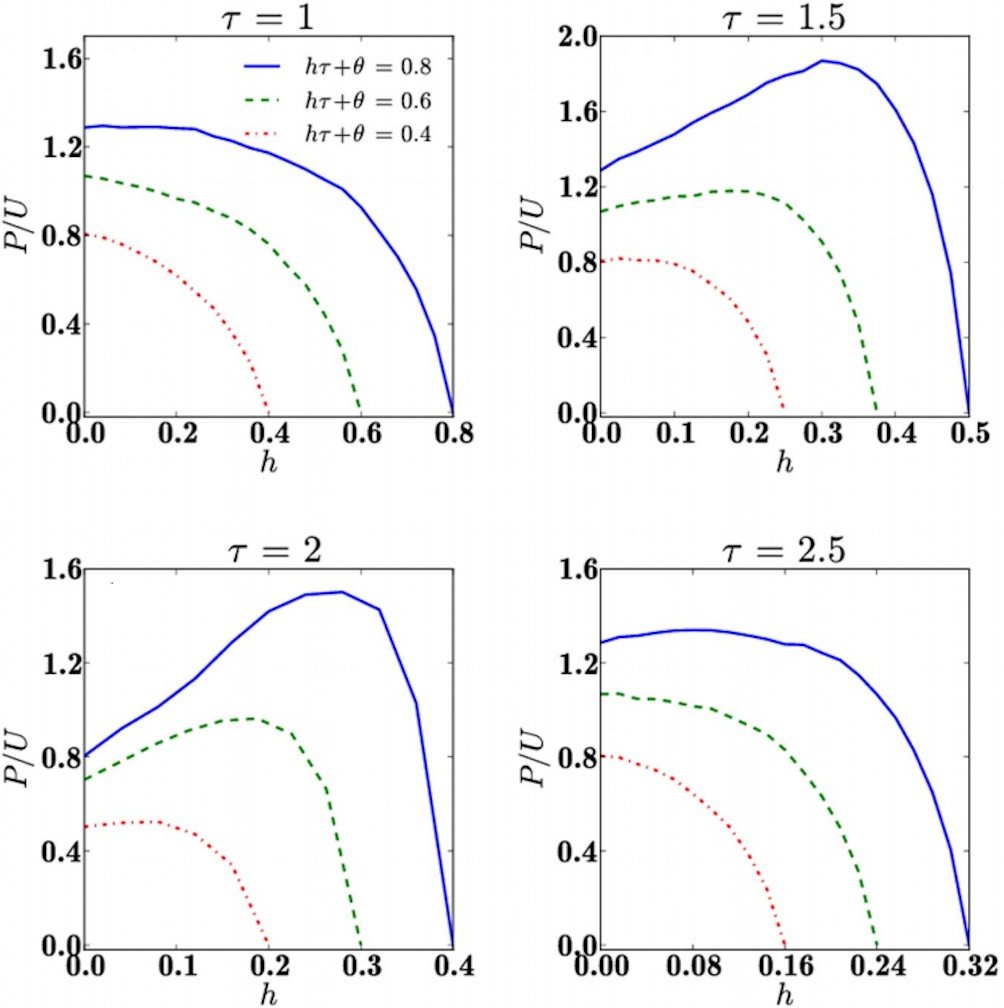}}
\caption{Minimizing recidivism requires carefully balanced rehabilitation programs, where both punishment and reward play a crucial role. Either neglecting punishment in favor of generous rehabilitation or vice versa will ultimately fail in successfully reintegrating offenders into society. Depicted is the ratio between paladins and unreformables $P/U$ in dependence on the amount of resources for rehabilitation $h$, as obtained for different values of the duration of intervention $\tau$ (see top of individual graphs). In all cases the severity of punishment $\theta$ is adjusted so that $h\tau + \theta = C$ (see legend in the top left graph), taking into account the fact that available resources are finite. The upper right graph reveals that the optimal parameter values are $h=0.3$, $\tau=1.5$ and $\theta=0.35$, which indicates that the most successful strategy is to allocate the limited resources so that after being punished, criminals experience impactful intervention programs, especially during the first stages of their return to society. For further details we refer to \cite{BERENJI_PONE14}, from where this figure has been reproduced.}
\label{optimal}
\end{figure}

While the majority of previous studies addressing the ``stick versus carrot'' dilemma concluded that peer punishment is more effective than reward in sustaining cooperation \cite{SIGMUND_PNAS01, SIGMUND_TREE07}, evidence suggesting that rewards may be as effective as peer punishment and lead to higher total earnings without potential damage to reputation \cite{MILINSKI_N02} or fear from retaliation \cite{DREBER_N08} is mounting rapidly. Moreover, in their recent paper \cite{RAND_NC11}, Rand and Nowak provide firm evidence that antisocial punishment renders the concept of sanctioning ineffective, and argue further that healthy levels of cooperation are likelier to be achieved through less destructive means. Regardless of whether the burden of cooperation promotion is placed on peer punishment \cite{GACHTER_S08, BOYD_S10, PERC_NJP12} or reward \cite{HILBE_PRSB10, SZOLNOKI_EPL10, SZOLNOKI_NJP12}, the problem with both actions is that they are costly. Cooperators who abstain from either punishing or rewarding therefore become ``second-order free-riders'', and they can seriously challenge the success of sanctioning \cite{PANCHANATHAN_N04, FEHR_N04} as well as rewarding \cite{SZOLNOKI_EPL10}. In the context of rehabilitating criminals, the question is how much punishment for the crime and how much reward for eschewing wrongdoing in the future is in order for optimal results, as well as whether these efforts should be placed on individuals or institutions \cite{sigmund_n10, szolnoki_pre11, szolnoki_prx13}, and assuming of course the resources are limited \cite{perc_srep12, chen_xj_f14}.

Berenji et al.~\cite{BERENJI_PONE14} have introduced an evolutionary game to study the effects of ``carrot and stick'' intervention programs on criminal recidivism. The model assumes that each player may commit crimes and may be arrested after a criminal offense. In the case of a conviction, a criminal is punished and later given resources for rehabilitation, in order to prevent recidivism. After their release into society, players may choose to continue committing crimes or to become paladins ($P$), implying they have been permanently reformed. Players are given $r$ chances to become paladins; if after the $r$-th arrest and rehabilitation phase, an individual relapses into crime, he or she is marked as an ureformable ($U$). States $P$ and $U$ are thus sinks, meaning they represent the end of the evolutionary process for each particular individual. As such the final $P/U$ ratio is a natural order parameter of the system: crime-infested societies are marked by $P/U \to 0$ and crime-free societies by $P/U \to \infty$. The main parameters of the game are the allocated resources for rehabilitation $h$, the duration of the rehabilitation intervention $\tau$, and the severity of punishment $\theta$, The parametrization of the game requires that for each player a record is kept for the number of punished and unpunished crimes. Stochastic simulations are performed which include the constraint $h \tau + \theta = C$, where $C$ is the total amount of available resources. Here $h \tau$ is the portion of the resources spent on rehabilitation efforts -- the carrots -- while $\theta$ is  the remainder, spent on punishment efforts -- the sticks. Because $C$ is finite, increasing one effort decreases the other, hence the ``stick versus carrot'' dilemma. For a given set of resource allocation $h,\tau,\theta$, we use the $P/U$ ratio as a measure of success. Figure~\ref{optimal} shows that as $C$ increases, the ratio $P/U$ will increase as well: with more general resources available, the conversion to paladins becomes more efficient. For a given value of $C$ Fig.~\ref{optimal} also shows that the most successful strategy in reducing crime, warranting the highest $P/U$ ratio, is to optimally allocate resources so that after being punished, criminals experience impactful intervention programs, especially during the first stages of their return to society. Indeed, the upper right panel of Fig.~\ref{optimal} reveals that for the case of $N=400$ players the optimal parameter values are $h=0.3$, $\tau=1.5$ and $\theta=0.35$. This indicates that the available resources $C$ need to be balanced so that there is enough stick (a sufficiently high $\theta$) and enough carrots (a sufficiently high $h$)  for a long enough time (a sufficiently high $\tau$). Within this model, excessively harsh or lenient punishments are less effective than the judicious balancing of the two. In the first case, there are not enough resources for rehabilitation left, in the second, punishment was not strong enough to discourage criminals from committing further crimes upon release to society. These findings have important sociological implications, and they provide clear guidance on how to minimize recidivism while maximizing social reintegration of criminal offenders.

\section{Summary and outlook}
\label{sumlook}
As we hope this review shows, the statistical physics of crime can provide useful insights into the emergence of criminal behavior, as well as suggest effective policies towards crime abatement. The mathematical model for crime hotspots reviewed in Section~\ref{hotspots}, for example, provides a mechanistic explanation for recent difficulties in observing crime displacement in experimental field tests of hotspot policing. The model also forms the basis for a better understanding of why and how crime hotspots form and propagate through time and space. Moreover, the position of the criminals and the biasing attractiveness field create nonlinear feedback loops, which give rise to complex patterns of aggregation that are reminiscent of actual crime hotspots.

In Section~\ref{point}, we reviewed how the highly space-time clustered nature of certain types of crime, akin to earthquakes and their aftershocks, can be exploited by means of self-exciting point process modeling. Methods developed in the realm of self-exciting point processes are well suited for criminological applications, and they have been applied successfully for gaining insight into the form of space-time triggering and temporal trends in urban crime, for geographical profiling of criminal behavior, as well as for modeling the temporal dynamics of violence in Iraq.

If crime is treated as a social dilemma, as reviewed in Section~\ref{social}, evolutionary dynamics reveals that informants are key to the emergence of a crime-free society. Furthermore, even a crime-dominated society can be transitioned to one that is largely crime-free by introducing an optimal number of informants. Since resources for their recruitment may be limited, an optimal control problem can be designed to find the most favorable informant recruitment strategies under different constraints. Human experiments fully confirm that informants are vital in diminishing crime, in fact even more so than predicted by the accompanying theory. Another evolutionary game designed to study crime, the conceptually simple three-strategy inspection game, reveals surprisingly nuanced and rich outcomes, including recurrent behaviors when there are gains associated with committing crimes. The complex dynamics that emerges from both games highlight that crime may be only partially understood by assuming that particular individuals are marked by a ``criminal nature''. Rather one should look at the overall social context and conditions that seem to promote criminal behavior.

In Section~\ref{gangs} we reviewed a recent study on cannabis production networks in the Netherlands, showing that all possible attempts at network disruption did not weaken the network as desired, but rather made it more resilient. This highlights the difficulties of policymakers and law enforcement agencies across the globe to find effective strategies to control and efficiently dismantle criminal networks. We have also reviewed two distinct attempts to identify criminal and gang networks using field and/or gang rivalry data, geographical information, social interactions and behavioral patterns. Both methods were successful in reconstructing known crime networks and gang clustering, showing how these case studies can be applied to situations where actual gang or network structures are not known and as possible guidance when intervening in trying to detect the source of crime. A more mathematical approach was used to model gang aggregation and territory formation by means of an Hamiltonian, Potts-like model, where interactions among agents were expressed via indirect graffiti markings. The analysis reveals that first and second order phase transitions from coexisting, well mixed gangs towards networked, geographically segregated gang clusters are possible even in the absence of direct gang-to-gang interactions. Under certain conditions, the indirect coupling provided by graffiti marking is thus sufficient to nucleate exclusive gang turfs.

Lastly, in Section~\ref{rehab}, we focused on rehabilitation and recidivism as modeled via a ``stick versus carrot'' evolutionary game. Given that total resources are finite, an important question to address is how much punishment for the crime and how much rehabilitation efforts after the punishment phase are in order to obtain optimal results. The reviewed research shows that the most successful strategy is a judicious resource allocation between the carrot and the stick, so that after sufficient punishment, criminals also experience impactful intervention programs. This is true especially during the first stages of a criminal's return to society. Excessively harsh punishments accompanied by too little rehabilitation, as well excessively lenient punishments appear to be not quite as effective in reducing the recidivism rate.

Extensions of approaches reviewed in Sections~\ref{hotspots}-\ref{rehab} may be useful to police and other security agencies in developing better and more cost-effective crime mitigation schemes while optimizing the use of their limited resources. The statistical physics of crime is still a very much developing and vibrant field, with ample opportunities for novel discoveries and improvements of existing models and theory. The model of crime hotspots, for example, could be easily upgraded to account for the distribution of real estate that better reflects the layout of an actual city. It would then be interesting to learn whether and how the introduced heterogeneity in the interaction network affects the emergence and diffusion of hotspots. If the crime is no longer residential burglary but crime that involves moving targets, further extensions towards social networks whose structure varies over time also become viable, and they point to a whole new class of coevolutionary crime models. If crime is treated as an evolutionary game the possibilities are even more, ranging from increased strategic complexity to the integration of more realistic, possibly coevolving, interaction networks that describe the societal fabric. In the realm of adversarial evolutionary games, it would also be interesting to study the impact of different strategy adoption rules, in particular since imitation-based rules are frequently contested with best-response dynamics in the realm of human behavior. In addition to the outlined extensions and upgrades of existing models, it is also possible to envisage new classes of models, especially such that would built more on self-organization and growth from first principles to eventually arrive at model societies with varying levels of crime. Here the hierarchical growth of criminal networks involving persuasion to join an organization and fidelity to either committing or not committing crimes appears to be a viable starting ground.

Informed by the reviewed research, we conclude that the statistical physics of crime clearly has far-reaching sociological implications, and we emphasize that the time is ripe for these insights to be used in synergy with traditional crime-related research to yield more effective crime mitigation policies. Many examples of ineffective policies clearly highlight that an insufficient understanding of the complex dynamical interactions underlying criminal activity may cause strong adverse effects of well-intended deterrence strategies. A new way of thinking, maybe even a new kind of science for deterring crime is thus needed -- in particular one that takes into account not just the obvious and similarly linear relations between various factors, but one that also looks particularly at the interdependence and interactions of each individual and its social environment. One then finds that this gives rise to strongly counterintuitive results that can only be understood as the outcome of emergent, collective dynamics, and this is why applied mathematics and methods of statistical physics can make important and substantial contributions to the understanding and containment of crime. We reiterate that the aim of this short review was to highlight these valuable theoretical resources that can help us bridge the widening gap between data and models of criminal activity, and we hope that the outlined directions for future research will further accelerate progress along this beautiful and highly applicable avenue of research.

\begin{acknowledgments}
This work was supported by the Army Research Office Multidisciplinary University Research Initiative grant W911NF-11-1-0332, the National Science Foundation grant DMS-1021850, the Slovenian Research Agency grant P5-0027, and by the Deanship of Scientific Research (DSR), King Abdulaziz
University, under grant 76-130-35-HiCi.
\end{acknowledgments}

\end{document}